\documentclass[pra,showpacs,twocolumn,groupaddress,superscriptaddress,nofootinbib,a4paper,accepted=2022-09-13,11pt]{quantumarticle}
\pdfoutput=1

\usepackage{times,amsmath,amsfonts,amssymb,latexsym}
\usepackage{graphicx,epsf}
\usepackage{subfigure}
\usepackage[shortlabels]{enumitem}
\usepackage{mathtools}
\usepackage{color}
\usepackage{soul}
\usepackage{algorithm}
\usepackage{algorithmic}
\usepackage{float}
\usepackage{amsmath}
\usepackage{bm}
\newcommand{\be}{\begin{equation}}
\newcommand{\ee}{\end{equation}}
\newcommand{\ba}{\begin{eqnarray}}
\newcommand{\ea}{\end{eqnarray}}

\newcommand{\ignore}[1]{}

\newcommand{\ket}[1]{\left | {#1} \right \rangle }

\newcommand{\bra}[1]{\left \langle {#1} \right | }

\def\CC{{\rm\kern.24em \vrule width.04em height1.46ex depth-.07ex
    \kern-.30em C}}
\def\P{{\rm I\kern-.25em P}}
\def\RR{{\rm
         \vrule width.04em height1.58ex depth-.0ex
         \kern-.04em R}}

\def\bbbc{{\mathchoice {\setbox0=\hbox{$\displaystyle\rm C$}\hbox{\hbox
to0pt{\kern0.4\wd0\vrule height0.9\ht0\hss}\box0}}
{\setbox0=\hbox{$\textstyle\rm C$}\hbox{\hbox
to0pt{\kern0.4\wd0\vrule height0.9\ht0\hss}\box0}}
{\setbox0=\hbox{$\scriptstyle\rm C$}\hbox{\hbox
to0pt{\kern0.4\wd0\vrule height0.9\ht0\hss}\box0}}
{\setbox0=\hbox{$\scriptscriptstyle\rm C$}\hbox{\hbox
to0pt{\kern0.4\wd0\vrule height0.9\ht0\hss}\box0}}}}
\def\bbbz{{\mathchoice {\hbox{$\sf\textstyle Z\kern-0.4em Z$}}
{\hbox{$\sf\textstyle Z\kern-0.4em Z$}}
{\hbox{$\sf\scriptstyle Z\kern-0.3em Z$}}
{\hbox{$\sf\scriptscriptstyle Z\kern-0.2em Z$}}}}

\usepackage{hyperref}

\begin{document}

\title{Transitions in Entanglement Complexity in Random Circuits}

\author{Sarah True}
\affiliation{Physics Department,  University of Massachusetts Boston,  02125, USA}
%\affiliation{Univ. Grenoble Alpes, CNRS, LPMMC, 38000 Grenoble, France}

\author{Alioscia Hamma}
\affiliation{Physics Department,  University of Massachusetts Boston,  02125, USA}
\affiliation{Dipartimento di Fisica `Ettore Pancini', Universit\`a degli Studi di Napoli Federico II, Via Cintia 80126,  Napoli, Italy}
\affiliation{INFN, Sezione di Napoli, Italy}

\begin{abstract}
Entanglement is the defining characteristic of quantum mechanics. Bipartite entanglement is characterized by the von Neumann entropy. Entanglement is not just described by a number, however; it is also characterized by its level of complexity. The complexity of entanglement is at the root of the onset of quantum chaos, universal distribution of entanglement spectrum statistics, hardness of a disentangling algorithm and of the quantum machine learning of an unknown random circuit, and universal temporal entanglement fluctuations. In this paper, we numerically show how a crossover from a simple pattern of entanglement to a universal, complex pattern can be driven by doping a random Clifford circuit with $T$ gates. This work shows that quantum complexity and complex entanglement stem from the conjunction of entanglement and non-stabilizer resources, also known as magic.

\end{abstract}

% \pacs{}
\maketitle

%%%%%%%%%%%%%%%%%%%%%%%%%%%%%%%%%%%%%%%%%%%%%%%%%
%SECTION INTRODUCTION
%%%%%%%%%%%%%%%%%%%%%%%%%%%%%%%%%%%%%%%%%%%%%%%%%
\section{Introduction}

One of the most fundamental properties of nature is the inevitable creation of disorder from order, leading to equilibration and thermalization. Classically, such behavior arises from extreme sensitivity to initial conditions \cite{eckmann1985ergodic, Rickles933, boeing2016visual, strogatz:2015}, that is, {\em chaos}. High sensitivity to initial conditions, also dubbed the ``butterfly effect'',  is easily observed at the macroscopic level \cite{strogatz:2015, boeing2016visual}.  In quantum mechanics, the characterization of chaos in quantum systems is an ongoing enterprise that is central in the field of quantum information and quantum many-body theory \cite{haake1991quantum, cotler2018out, bhattacharyya2019web, chaudhury2009quantum, roberts2017chaos, roberts2016bound}. For example, studies in random matrix theory \cite{haake1991quantum, atas2013distribution, cotler2017chaos, cotler2017black, gharibyan2018onset, Oliviero2020random, leone2020isospectral, rao2020wigner} have provided analytical means of identifying chaos in quantum Hamiltonians and systems of entangled qubits. Indeed, it is well known that the emergence of chaos in quantum systems goes hand in hand with entanglement \cite{Wang2004entanglement, chen2018universal, hosur2016chaos, liu2018entanglement, kumari2019untangling, roberts2017chaos, Hamma_2012, Hammalungo_2012}, a fact that has brought the problem of creating sufficiently complex entanglement to the forefront of quantum computing \cite{jozsa1997entanglement, preskill2012quantum} as physicists endeavour to recreate the information scrambling properties of nature's most chaotic systems such as black holes \cite{sekino2008fast, hayden2007black, Landsman2019verified, cotler2017black, preskill2012quantum, yoshida2017efficient, ding2016conditional, roberts2016bound, swingle2016measuring}.

In a quantum computer, qubits are entangled by a circuit of unitary logic gates. Clifford gates represent an important subgroup of the unitary group because they can be efficiently simulated by a classical computer, and yet are still very effective in attaining maximum entanglement \cite{Gottesman:1998hu, Nielsen, roberts2017chaos, harrow2017quantum}. However, while these gates are easily simulated on classical machines, they are only capable of producing non-complex patterns of entanglement, and are therefore not sufficient for true chaotic behavior in quantum systems \cite{Gottesman:1998hu, feynman1982simulating, roberts2017chaos, leone2021quantum, oliviero2021transitions, bravyi2016improved, gross2020quantum, harrow2017quantum}. Including gates outside the Clifford group provides a set that is universal for quantum computation and induces complex entanglement and chaotic behavior \cite{BOYKIN2000101, Nielsen, gottesman2009introduction, Hamma_2012, Hammalungo_2012}. The expensive nature of such gates introduces yet another problem, that of optimizing the use of non-Clifford resources \cite{ross2016optimal, litinski2019, 2019, wang2021, gheorghiu2021tcount, leone2021quantum, oliviero2021transitions, bravyi2016improved, gross2020quantum}. As such, a greater understanding of what lies between Clifford and universal entanglement is necessary not only in the development of quantum technology, but also in painting a complete picture of the transition in quantum physics from order to chaos. We therefore aim to study quantum circuits that induce different levels of complexity between these two extremes.

\begin{figure*}
    \centering
    \includegraphics[height=8cm]{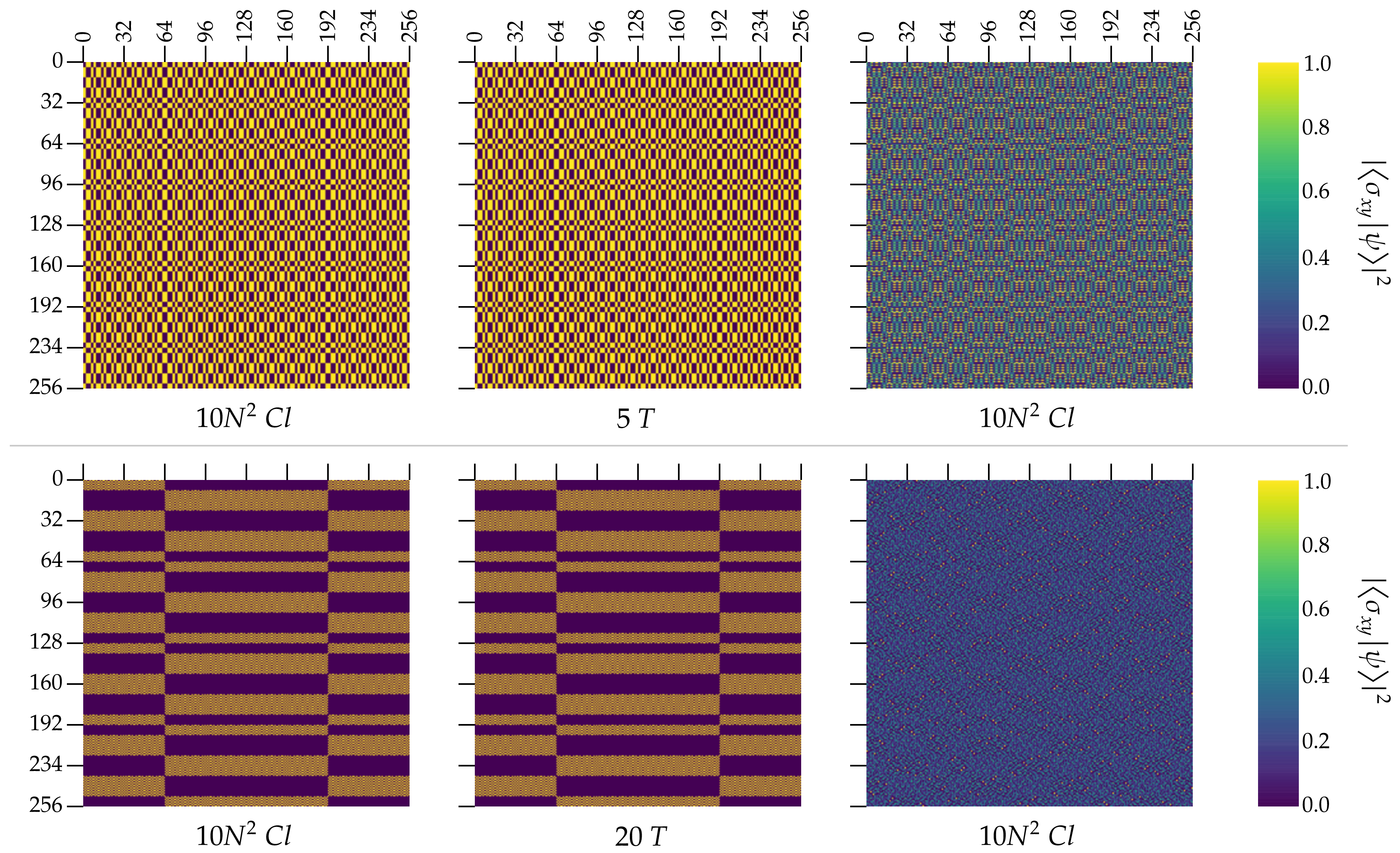}
    \caption{Color map representation of a 16-qubit state after $10N^2$ random Clifford gates, then $n_T$ random $T$ gates $($\textit{Top row:} $n_T{=}5$. \textit{Bottom row:} $n_T{=}20)$, then another $10N^2$ random Clifford gates are applied, starting with the initial state $\ket{\psi_0}{=}\ket{0}^{\otimes N}$. The image is arranged by partitioning the state into two equal subsystems and expanding one along each axis. A pixel at some position $(x, y)$ is mapped from the magnitude of the amplitude for the computational basis state $\ket{\sigma_{xy}}{=}\ket{\sigma_x}{\otimes}\ket{\sigma_y}$ $($for example, the pixel at $(0, 0)$ corresponds to the state $\ket{\sigma_{00}}{=}\ket{0}^{\otimes N/2}{\otimes}\ket{0}^{\otimes N/2})$. %While neither layer of $T$ gates shows any discernible pattern change immediately after being applied, the
    The action of a $T$ gate, being a phase gate, does not change the color map representation at all. However, after being spread around by the second Clifford circuit, the color map representation becomes  more mixed for the circuit with the larger number of inserted $T$ gates, showing the interplay between the non-Clifford resources and the operator spreading through Clifford-driven entanglement in the onset of quantum chaos. 
    }
    \label{scrambling}
\end{figure*}

In this paper, we study the onset of complex entanglement in systems of qubits by numerically simulating random Clifford circuits doped with a controlled number $n_T$ of $T$ gates -- which do not belong to the Clifford group -- showing the full transition from non-chaotic to chaotic behavior and from non-complex to complex entanglement by the emergence of typical entanglement properties belonging to random universal circuits. In particular, we monitor (i) the entanglement spectrum statistics, showing that, as the $T-$doping increases, we obtain a crossover towards the universal Wigner-Dyson distribution for the gaps in the reduced density matrix spectrum, the so-called Entanglement Spectrum Statistics (ESS) \cite{atas2013distribution, chamon2014emergent, shaffer2014irreversibility, zhou2020single, yang2017entanglement}. We also compute (ii) the fluctuations of entanglement, showing how doping drives the transition towards universal fluctuations computed through the Haar measure on the unitary group \cite{Hamma_2012, leone2021quantum, oliviero2021transitions, Hammalungo_2012}. Finally, we demonstrate (iii) that the quantum complexity conjured by entanglement and $T$ gates is characterized by the hardness of the entanglement cooling algorithm and find a quantitative relation between the doping $n_T$ and the performance of the algorithm in undoing any initial scrambling \cite{shaffer2014irreversibility, chamon2014emergent}. As the disentangling algorithm allows the learning of a random quantum circuit, these results show how the quantum machine learning of a random circuit increases in complexity with the number of $T$ gates.

%%%%%%%%%%%%%%%%%%%%%%%%%%%%%%%%%%%%%%%%%%%%%%%%%
%TECHNICAL
%%%%%%%%%%%%%%%%%%%%%%%%%%%%%%%%%%%%%%%%%%%%%%%%%

\section{Complexity as Revealed by the Entanglement Spectrum}

\subsection{Entanglement Heating Setup}\label{heating}
In this section, we describe the setup of the problem. Each simulation begins with a process called {\em entanglement heating}, during which a random quantum circuit is used to produce a highly entangled state. We begin with a system of $N$ qubits initialized in the zero state of the computational basis $\ket{\psi_0}{=}\ket{0}^{\otimes N}$. The dimension of the Hilbert space will be denoted by $d{=}2^N$. The entanglement heating process involves applying a random unitary circuit $U$ containing an assortment of $10N^2$ quantum  gates chosen from a predetermined set, either Clifford $\left(U^0\right)$ or universal $\left(U^U\right)$. The Clifford set includes the $C\! N\! O \! T$, $H\! adamard$, and $S$ gates $\left(P\left(\frac{\pi}{2}\right)\right)$. A universal gate set is created by including $T$ gates $\left(P\left(\frac{\pi}{4}\right)\right)$ along with the Clifford set \cite{Gottesman:1998hu}. A doped Clifford${+}T$ circuit $\left(U^{n_T}\right)$ is a random Clifford circuit in which a controlled number $n_T$ of non-Clifford resources -- in this case, $T$ gates -- are inserted. The actions of $\pi/4$-phase shifts ($T$ gates) alone have no effect on entanglement, as single qubit operations do not change the Schmidt decomposition of a state. However, when utilized in conjunction with the Clifford set (which includes the two-qubit $C\! N\! O\! T$ gate), a $T$ gate applied to some qubit acts as a \textit{seed} of chaos. These seeds are subsequently propagated by random Clifford gates, resulting in operator spreading and growth \cite{nahum2017quantum, nahum2018operator, zhou2020single, mi2021information}, therefore unleashing chaotic behavior \cite{roberts2015shocks, moudgalya2019operator}. This effect can be seen in Fig. \ref{scrambling}, where a visualization of the bipartite weights of the wave function is shown at three different stages of a circuit: first after a random Clifford circuit, then after a layer of $T$ gates, and finally after a second Clifford circuit. As we can see, the complexity resulting from operator spreading is evident already at the visual level, which suggests a machine-learning approach to the detection of quantum chaos \cite{zhou2020single}.

\begin{figure}[h!]
    \centering
    \includegraphics[width=\linewidth]{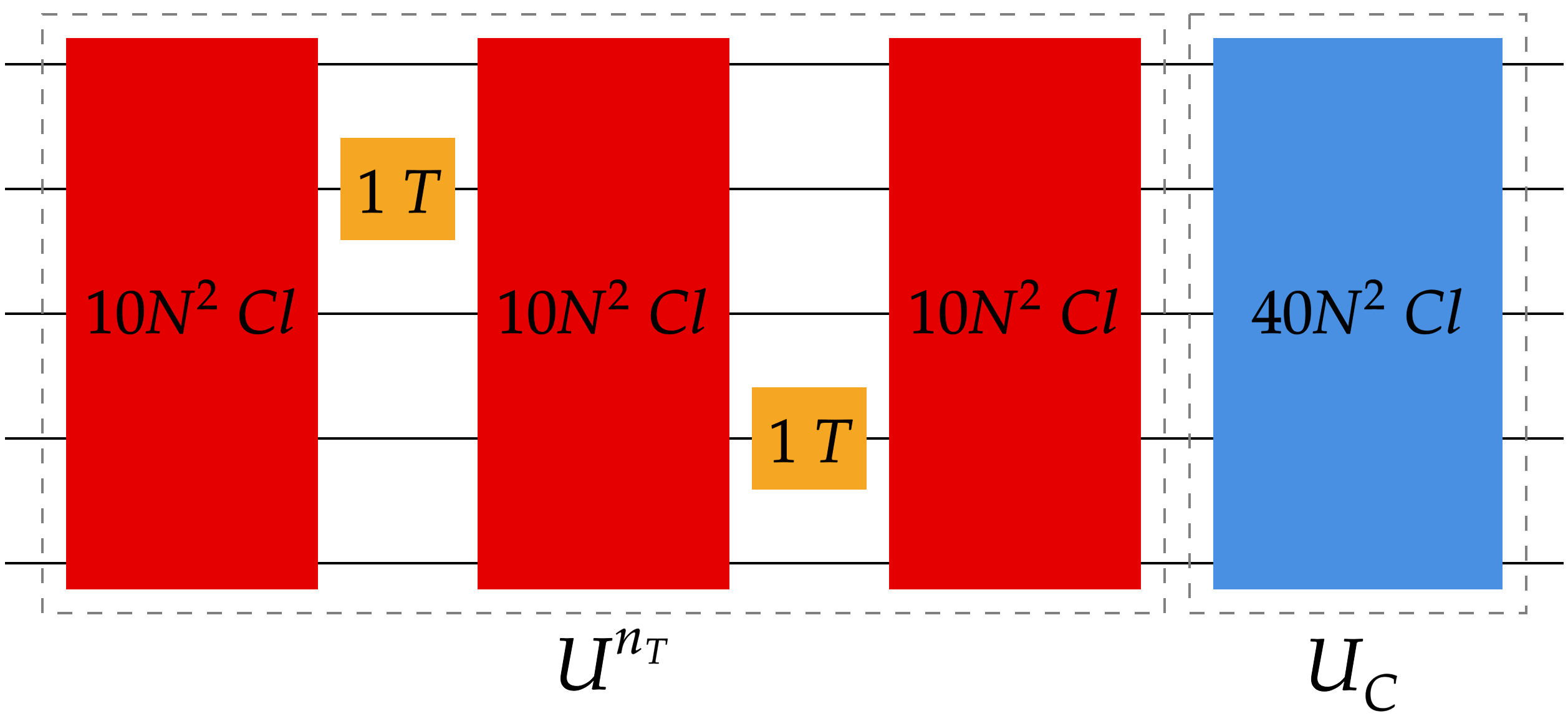}
    \caption{Schematic of the Clifford${+}T$ circuit design used for entanglement heating ($U^{n_T}$) and entanglement cooling ($U_C$). In the heating circuit, single $T$ gates are inserted between blocks of $10N^2$ Clifford gates, thus $n_T$ is also the number of $T-$layers. For example, $n_T{=}2$ in the circuit $U^2$ shown here. Regardless of the number of layers in the heating circuit, the subsequent cooling circuit always contains a single block of $40N^2$ Clifford gates (the exception to this is the case of infinite temperature cooling performed in section \ref{fluctuations}, during which $10N^2$ gates are used).}
    \label{circuit}
\end{figure}

In the following simulations, we employ Clifford${+}T$ circuits $U^{n_T}$ designed by alternating single $T$ gates and operator-spreading blocks of $10N^2$ Clifford gates for a total of $n_T$ $T-$layers between $n_T{+}1$ Clifford blocks, as shown in Fig. \ref{circuit}. The choice of $T$ gate placement (i.e. which qubit it acts on) is not a contributing factor during this process \cite{leone2021quantum}, and so is chosen at random. We perform exact numerical simulations of these circuits ranging from $N{=}8$ to $N{=}18$ qubits, with various values of $n_T{\in}\left[0,40\right]$. For every doping $n_T$, $200$ realizations are run to obtain an ensemble of states $\{U_i^{n_T}\ket{\psi_0}\}_{i=1}^{200}$. In addition, for each instance of $N$, we generate states $\{U_i^U\ket{\psi_0}\}_{i=1}^{200}$ evolved with random universal circuits $U^U$.

\subsection{Post-Heating Level Spacing Statistics}

\begin{figure}[h!]
    \centering
    \includegraphics[width=\linewidth]{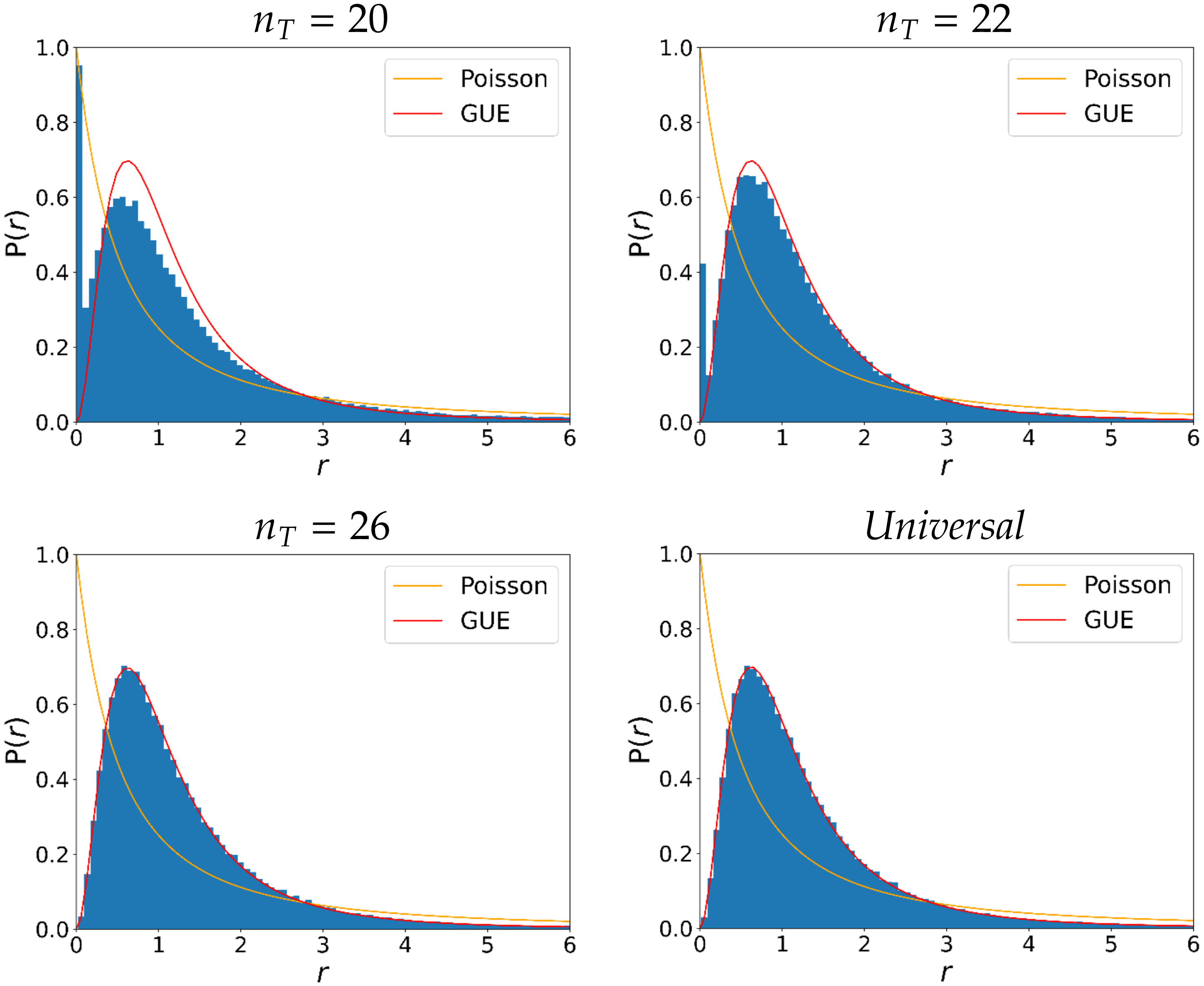}
    \caption{Gap distribution in the entanglement eigenvalues (ESS) for Clifford${+}T$ circuits with different doping $(n_T{=}20,22,26)$ applied to the initial zero state $\ket{\psi_0}$ for $N{=}18$ qubits, as well as an $18-$qubit universal circuit. Each plot is obtained with $200$ realizations. As the doping increases, a stricter adherence to the Wigner-Dyson distribution emerges.}
    \label{p(r)18}
\end{figure}

After heating, the output $\ket{\psi}{=}U^{n_T}\ket{\psi_0}$ of each circuit is analyzed  by characterizing entanglement complexity in terms of the ratio of adjacent gaps in the entanglement spectrum as defined in \cite{zhou2020single}. Given a composite state $\ket{\psi}{=}\ket{\psi_{A}}{\otimes}\ket{\psi_{B}}$ under equal bipartition such that each subsystem contains $N_A{=}N_B{=}N/2$ qubits, let $p_i{=}\lambda_i^2$ be the eigenvalues obtained from the Schmidt decomposition $\ket{\psi}{=}\sum_i\lambda_i\ket{i_A}\ket{i_B}$. The eigenvalues $\{p_k\}_A$ of the reduced density matrix $\rho_A{=}tr_B\left(\ket{\psi}\bra{\psi}\right)$ form the reduced density matrix spectrum. The level spacing ratios are calculated as $r_k{=}\left(p_{k-1}{-}p_k\right){/}\left(p_k{-}p_{k+1}\right)$ where the eigenvalues are ordered such that $p_k{>}p_{k+1}$. Any degeneracy is accounted for by including an additional instance of $r_k$ for each occurrence of $p_k$. In the case of a random circuit with universal gates (or just for a random unitary operator on the full space of $N$ qubits), one obtains the universal ESS $\{r_k\}$ exhibiting a Wigner-Dyson (W-D) distribution, following the surmise \cite{atas2013distribution} \be
P_{WD}(r){=}\left(r+r^2\right)^{\beta}{/}\left[Z\left(1+r+r^2\right)^{1+3\beta/2}\right]
\ee
where the Gaussian Unitary Ensemble $P_{GUE}(r)$ is given by $Z{=}4\pi/81\sqrt{3}$ and $\beta{=}2$. Given this known property, one way we characterize a state's entanglement complexity is by the degree of similarity between the distribution $P(r)$ of its entanglement spectrum and that of a Gaussian $P_{GUE}(r)$. We quantify this statistical distance by calculating the Kullback-Leibler divergence:
\be
D_{KL}\left[P(r)||P_{GUE}(r)\right]{=}\sum_iP(r_i)\ln\frac{P(r_i)}{P_{GUE}(r_i)}.
\label{eqdkl}
\ee
Note that $D_{KL}\left[P(r)||P_{GUE}(r)\right]{=}0$ when $P(r){=}P_{GUE}(r)$.

%By performing this analysis on our $T$-doped circuits $U_i^{n_T}\ket{\psi_0}$, we observe more closely the emergence of a Gaussian distribution as non-Clifford resources are added. As expected, the states evolved with universal circuits $U^L\ket{\psi_0}$ exhibit an entanglement spectrum following a GUE distribution (Fig. \ref{p(r)18}. The same behavior arises as we increase the number of $T$ gates used to dope our Clifford circuits, as shown in Fig. \ref{p(r)18}. 

\begin{figure}[H]
    \centering
    \includegraphics[width=\linewidth]{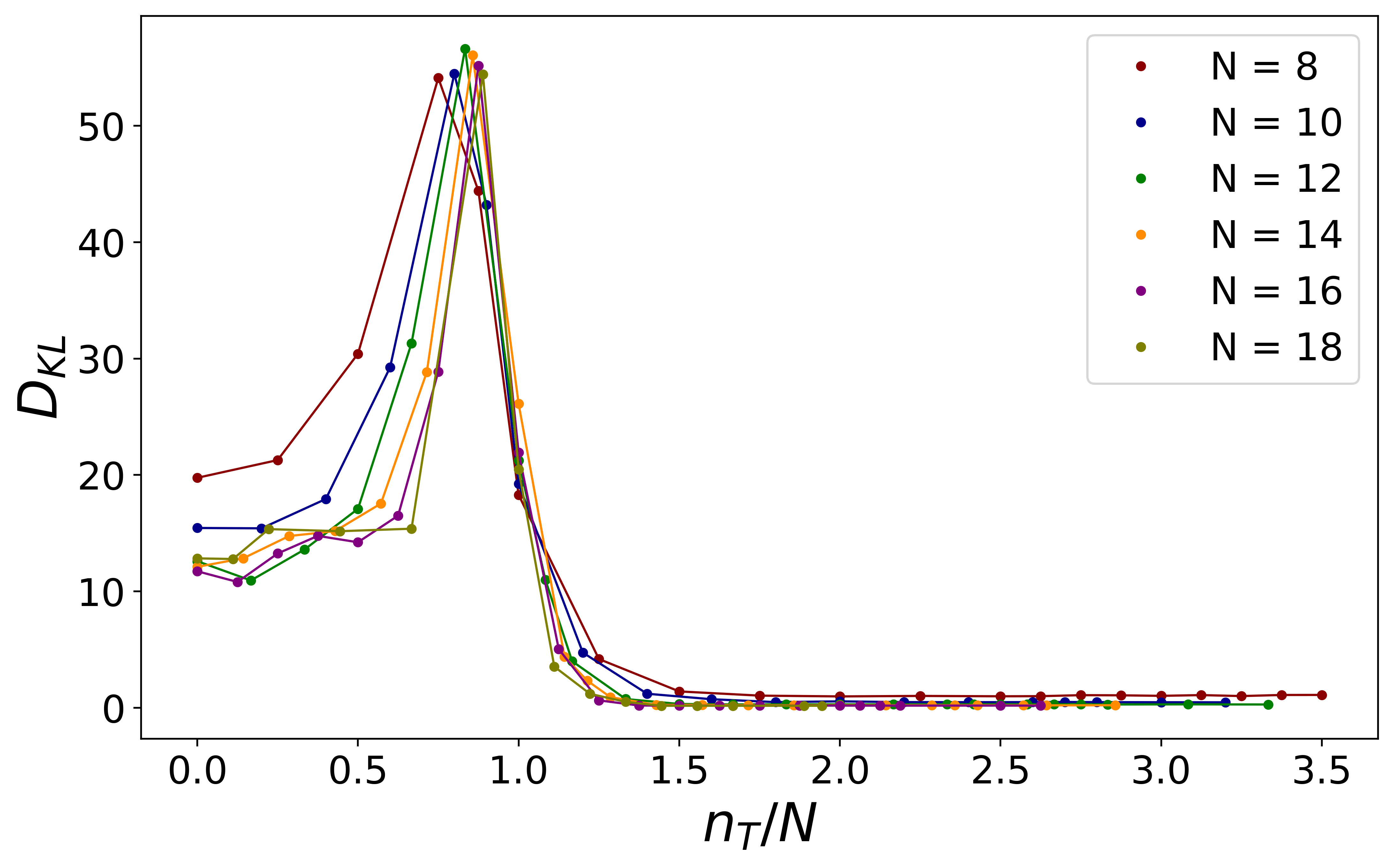}
    \caption{Statistical distance $D_{KL}$ (Eq. \eqref{eqdkl}) from W-D distribution plotted against $n_T/N$ for all system sizes. All curves collapse to a universal scaling function $f(n_T/N)$ for $n_t \geq N$ (Eq. \eqref{eq3.1.1}).}
    \label{dkl}
\end{figure}

As the doping level $n_T$ in the heating circuit increases, we observe the emergence of the Wigner-Dyson distribution for the ESS, see Fig. \ref{p(r)18}. The resulting calculations of Eq. \eqref{eqdkl} for each $(N,\ n_T)$ pair are shown in Fig. \ref{dkl}. We see that the divergence $D_{KL}$ displays a non-trivial behavior: as $n_T$ approaches $N$, there is an extensive divergence $D_{KL}$. This is the signature of a sudden change in the pattern of entanglement, which can occur despite the absence of other typical critical behaviors (e.g. the change between different patterns of entanglement across the factorization point in spin chains \cite{amico2006}). Above this critical value, we can fit the data for $n_T{\geq}N$ with the formula
%In addition to visual confirmation of this effect, the impact that T-doping has on these statistics is summarized by an analysis of the $D_{KL}$ (Eq. \ref{eq2.2.2}) when calculated for all Clifford + T circuits such that $n_T \geq N$:
\ba
D_{KL} &=& 24\left[ d^{1.25(1-n_T/N)} + \frac{1}{d^{0.6}} \right] + 0.16
\label{eq3.1.1}\\
\nonumber
&\equiv&  D_{KL}^{uni}+\epsilon (n_T)
\ea
For large $n_T$, the error $\epsilon\equiv 24 d^{1.25(1-n_T/N)}$ is exponentially suppressed and Eq. \eqref{eq3.1.1} yields the universal value $D_{KL}^{uni}=24/d^{0.6}+0.16$, which is in perfect agreement with the numerical value of $D_{KL}$ computed for a random universal quantum circuit. This result raises the question of how many $T$ gates one needs to add before the $D_{KL}$ crosses over and begins to approach the universal limit. In other words, we want to know the minimum value of $n_T$ such that the error is smaller than the universal value, that is,
\be
\epsilon ({n_T}) < \frac{24}{d^{0.6}} + 0.16.
\label{eq3.1.2}
\ee
This minimum $n_T^{min_D}$ is calculated for a given system size by plugging $\epsilon ({n_T})\equiv 24d^{1.25(1-n_T/N)}$ into Eq. \eqref{eq3.1.2} and solving for $n_T$, rounding up to the nearest integer. In other words, given Eq. \eqref{eq3.1.2} rearranged in the form $n_T > f(N)$, we define $n_T^{min_D}(N)\coloneqq\lceil f(N)\rceil$. We perform this calculation for all $N\in\left[0, 40\right]$ and fit the results, finding that
\be
n_T^{min_D} = N+2.
\label{eq3.1.3}
\ee

This result, in addition to the collapse shown in Fig. \ref{dkl}, demonstrates a linear scaling with $N$ of the number of $T$ gates $n_T$ needed to attain the universal Wigner-Dyson distribution. In other words, a finite density of $T$ gates drives the transition towards universal ESS. This result is to be compared with that of \cite{zhou2020single}, where it was shown that a single $T$ gate on a {\em random} product state was sufficient to obtain the transition. However, a random initial product state contains an $\mathcal{O}(N)$ number of non-Clifford resources in the preparation of the initial state. It is remarkable, though, that one seed of operator spreading is both necessary and sufficient for the transition after initial state preparation with a random product state.

\section{Fluctuations of Entanglement Entropy}\label{fluctuations}

%% \cite{liu2018entanglement}
In this section, we study the temporal entanglement fluctuations under a random Clifford circuit following the entanglement heating process, schematized with a blue block in Fig.\ref{circuit}, the so-called cooling circuit. If the Clifford circuit is completely random, this corresponds to an infinite temperature of the cooling algorithm, see section \ref{coolingalg}.

We monitor the entanglement in the system during both the heating and the cooling procedure as a function of time, that is, after every new gate in the circuit. 
%The changes in entanglement occurring within the system are tracked by performing the following calculations after each gate application. 
The  entanglement between certain qubits at a particular point in time is determined again by partitioning the state into two subsystems $\ket{\psi}{=}\ket{\psi_A}{\otimes}\ket{\psi_B}$ of dimensions $d_A{=}2^{N_A}$ and $d_B{=}2^{N_B}$ such that $N{=}N_A{+}N_B$ (note that it is no longer necessary that $N_A{=}N_B$). We then calculate the Von Neumann entropy of $\ket{\psi_A}$ by
$S\left(\psi_A\right){=}{-}\sum_{i=1}^{d_A}p_i\log p_i$.
%In the case of a completely factorized state (such as our initial state $\ket{\psi_0}$), $S\left(\psi_A\right){=}0$ for any choice of $\ket{\psi_A}$.
In order to fully consider the relative entanglement between various subsystems, this entropy is calculated for all ${N}{-}{1}$ possible bipartitions and averaged to obtain a final value $\overline{S}_k$, with $k$ representing the number of gates that have so far been applied. 
With this definition, $\overline{S}_k=0$ if and only if the state is completely factorized. At the end of the circuit $U$, all recorded $\overline{S}_k$ form a set $S^U\coloneqq \{\overline{S}_k\}_U$. By performing these calculations after each gate, we observe the inevitable increase in entanglement entropy that has come to be expected under random unitary transformations, as well as its eventual saturation at a maximum value \cite{shaffer2014irreversibility, Hamma_2012, popescu2009evolution, Hammalungo_2012}. 

After the entanglement heating, the system has typically reached an equilibrium, with a nearly maximal value for the entanglement entropy. If one then continues to apply random Clifford operations, no further drift will be observed; rather, only fluctuations around the equilibrium value will occur. We analyze these fluctuations through further stochastic evolution of the fully thermalized state $U\ket{\psi_0}$ with a new random circuit $U_C$ of $10N^2$ Clifford gates (this is an example of the cooling algorithm at infinite temperature, see section \ref{coolingalg}). We continue to calculate the entanglement entropy $S(\psi_A)$ at each step in the circuit, giving a new set of quantities $S^{U_C}$ \textit{in addition to} our previously gathered data $S^U$. The temporal variance of $S^{U_C}$ is then calculated as
\be
Var^U \coloneqq Var(S^{U_C})=\left<\left(S^{U_C}-\left<S^{U_C}\right>\right)^2\right>
\label{eq2.3.1}
\ee
where $\langle\ldots\rangle$ denotes the temporal average of a single realization. We then average across all realizations for a final quantity $\overline{Var^U}$.

\begin{figure}[H]
    \centering
    \includegraphics[width=\linewidth]{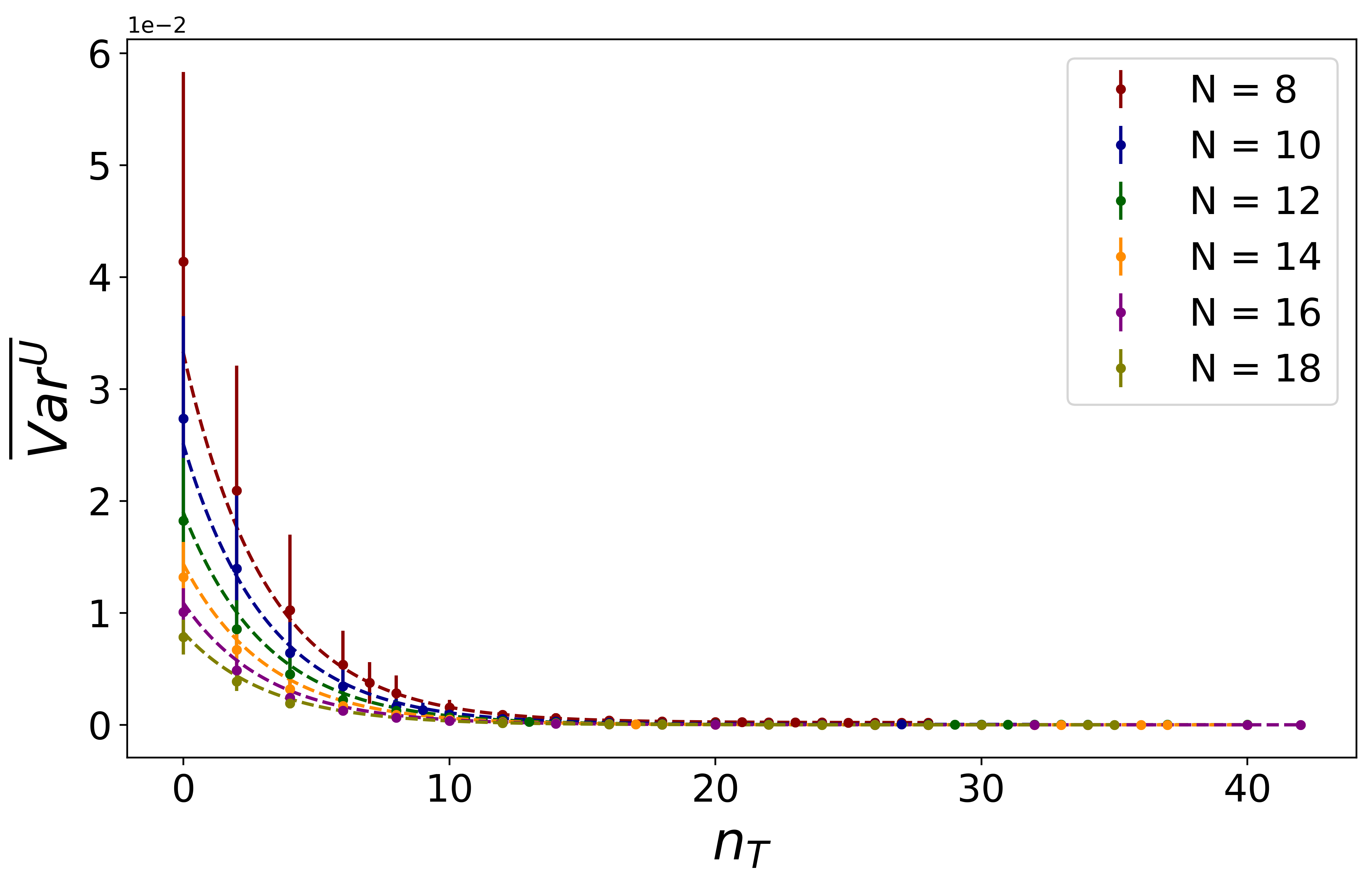}
    \caption{Sample average of the temporal variance $Var^U$ (Eq. \eqref{eq2.3.1}) for circuits $U_C$, which are applied to fully entangled states $U^{n_T}\ket{\psi_0}$ (see Fig. \ref{circuit}). The variance of $U_C$ can be seen trending to zero as the initial doping $n_T$ increases (fit by Eq. \eqref{eq3.2.1}).}
    \label{var}
\end{figure}

%%%%%%%%%%%
Studying the resulting variance calculations for all circuit realizations, we find a direct relationship between the average variance $\overline{Var^U}$ of entanglement entropy in $U_C$, the number of $T$ gates $n_T$ added to the heating circuit, and the dimension $d$ of the full system. In the case of a random universal circuit $U^U$ where $n_T$ is not a fixed quantity, a fit of the results shows that $\overline{Var^U}$ decreases with $d$ as $0.2/d^{1.25}$. The $T-$doped Clifford circuits likewise mimic this result for large values of $n_T$. Specifically, upon fitting the data (Fig. \ref{var}), we find that
\be
\overline{Var^U} = \frac{0.1}{d^{0.2}}\exp\left[\frac{-n_T}{3.15}\right] + \frac{0.2}{d^{1.25}},
\label{eq3.2.1}
\ee
demonstrating a definitive rate of change in entropy variance as more $T$ gates are added to the circuit and the system begins to approach universal behavior. As with the process of calculating Eqs. \eqref{eq3.1.2}-\eqref{eq3.1.3}, we obtain the minimum number of $T$ gates $n_T^{min_V}$ needed to approach the universal limit for each system size $\left(\right.$i.e. the minimum $n_T$ that satisfies the inequality $\frac{0.1}{d^{0.2}}\exp\left[\frac{-n_T}{3.15}\right] < \frac{0.2}{d^{1.25}}\left.\right)$, fitting the results to find
\be
n_T^{min_V} = 2.29N-\frac{5}{3}.
\label{eq3.2.2}
\ee
Comparing this result with Eq. \eqref{eq3.1.3}, we observe that the crossover to universal fluctuations occurs with the same linear scaling as  that of universal statistics. This result is in agreement with that of \cite{leone2021quantum, oliviero2021transitions}, where it was shown that universal fluctuations of subsystem purity in a doped random Clifford circuit also require a doping scaling with $N$. In other words, universal entanglement fluctuations are a good probe to the approach to universal entanglement spectrum statistics. 

\section{Doping, Complexity, and Entanglement Reversibility}\label{coolingalg}

We have seen that the entanglement entropy of any subsystem in $\ket{\psi}$ is statistically guaranteed to increase and saturate over time under a random quantum circuit, a phenomenon we called entanglement heating. If one knew everything about the heating circuit $U$, one could reverse the evolution and return to the initial factorized state $\ket{\psi_0}$ by simply acting with  $U^{\dag}$, bringing us to the state $U^{\dag}U\ket{\psi_0}{=}\ket{\psi_0}$. If, however, we instead only have access to the entangled state $U\ket{\psi_0}$ and not any information about the heating circuit, we must use other methods to find a new circuit that will undo the original one. Previous works \cite{chamon2014emergent, shaffer2014irreversibility} have shown that a Clifford circuit can be reversed by a Metropolis-like entanglement cooling algorithm. Unsurprisingly, though, a random quantum circuit realized with a universal set of gates cannot be reversed (except by exhaustive search). We ask the question of how the crossover between these two extreme behaviors is driven by the doping $n_T$.  If we gradually dope the circuit with one seed of chaos at a time, exactly how much is enough to make it virtually impossible to reverse? More specifically, if we dope the circuit with $\mathcal{O}(N)$ non-Clifford gates, thereby obtaining the transition to universal ESS and universal entanglement fluctuations, is this enough to make the entanglement pattern so complex that -- as in a completely random state -- no entanglement cooling is possible?

\subsection{Entanglement Cooling Setup}

Given the absence of an exact inverse circuit $U^{\dag}$, we implement a Metropolis-like algorithm to learn a sufficient entanglement cooling circuit $U_C$. This algorithm is described below as in Ref. \cite{shaffer2014irreversibility}:
\begin{algorithm}[H]
\caption{Entanglement Cooling Algorithm}\label{alg1}
\algsetup{indent=2em}
\begin{algorithmic}[1]
\STATE $S_{\mathrm{old}} \leftarrow$ entanglement entropy of $\psi$
\WHILE{$S_{\mathrm{old}}>0$}
\STATE Apply gate at random: $\psi \leftarrow \psi_{\mathrm{new}}$
\STATE $S_{new} \leftarrow$ entanglement entropy of $\psi$
\IF{$S_{\mathrm{new}}>S_{\mathrm{old}}$}
\STATE $r \leftarrow x \in[0, 1]$
\IF{$r>\mathrm{exp}\left[-\beta(S_{\mathrm{new}}-S_{\mathrm{old}})\right]$}
\STATE Undo gate: $\psi \leftarrow \psi_{\mathrm{old}}$
\ENDIF
\ENDIF
\STATE $S_{\mathrm{old}} \leftarrow S_{\mathrm{new}}$
\ENDWHILE
\end{algorithmic}
\end{algorithm}
 Starting with a maximally entangled state $U\ket{\psi_0}$, the way the algorithm builds $U_C$ is by applying a random gate, calculating the new entropy $S_{\mathrm{new}}$, and comparing it to the entropy of the previous state $S_{\mathrm{old}}$. In cases where $S_{\mathrm{new}}{>}S_{\mathrm{old}}$, the action is accepted with a probability ${\mathrm{exp}}{\left[-\beta(S_{\mathrm{new}}{-}S_{\mathrm{old}})\right]}$. In this algorithm, the abstract inverse temperature $\beta$ represents the probability of accepting a gate that increases the entropy. At infinite temperature $\beta{=}0$, every gate is accepted, resulting in an algorithm equivalent to the heating process outlined in section \ref{heating}. For the cooling algorithm, we use  $\beta{=}1\mathrm{e}4$. If a gate is rejected, we revert back to the previous state and try again. Any action that decreases the entropy is automatically accepted. This process is continued until either zero entropy is reached or a maximum of $40N^2$ gates have been applied (Fig. \ref{circuit}). In the former case, any remaining steps become ``do nothing'' operations, keeping the state constant until the $40N^2$ limit has been reached. In the end, we are left with a new set of entropy values $S^{U_C}$ (as with the fluctuation analysis, this set excludes the initial heating data $S^U$).
 
We now want to remark that the gates used to construct $U_C$ are chosen only from the Clifford group. Although the exact inverse of a $T-$doped circuit must include $T$ gates, allowing the cooling algorithm to choose non-Clifford operations ends up being counterproductive. The scrambling power of $T$ gates that features heavily during the heating process works against us when attempting to undo the entanglement, as we do not have the control necessary to direct that power exactly where we need it to go. We therefore work only with Clifford gates to improve the algorithm's chance of success.

When implemented to reverse a random Clifford circuit, this algorithm is known \cite{chamon2014emergent, shaffer2014irreversibility} to successfully disentangle a system of qubits. Because the entanglement in this case is very easy to undo, we characterize it as being \textit{non-complex}. If, however, $U$ is a random universal circuit, then the algorithm has no effect on entanglement, and as such $U_CU\ket{\psi_0}$ is just as entangled as $U\ket{\psi_0}$. In other words, $U\ket{\psi_0}$ has \textit{complex} entanglement \cite{shaffer2014irreversibility}. We quantify the algorithm's level of success in disentangling an $N-$qubit state $U\ket{\psi_0}$ by comparing the final entropy after cooling $S^{U_C}_f$ to $\overline{S^{U_{C,U}}_f}$, where $\overline{S^{U_{C,U}}_f}$ is the average final entropy over $200$ realizations of disentangling a random $N-$qubit universal circuit $U^U$. We define the \textit{reversibility} $R^U$ of the initial heating circuit $U$ using the  average (over the different realizations) difference per qubit between these two quantities: 
\be
R^U \coloneqq \frac{1}{N}\left<\overline{S^{U_{C,U}}_f}-S^{U_C}_f\right>
\label{eq2.4.1}
\ee
Given the algorithm's  ineffectiveness in disentangling Haar-random states -- that is, any $U^U\ket{\psi_0}$ obtained from a random circuit with universal gates \cite{shaffer2014irreversibility} -- a relatively large $R^U$ would indicate that the complexity of entanglement present in $U\ket{\psi_0}$ is significantly lower than that of the universal case. On the other hand, complex entanglement in $U\ket{\psi_0}$ would cause $R^U$ to approach zero as the entanglement becomes just as hard for the algorithm to reverse as that generated by $U^U$. As noted in \cite{zhou2020single}, factorizing the state is equivalent to learning one column of the random circuit. The complexity of entanglement due to the spreading of non-Clifford resources is thus also connected to the hardness of the quantum machine learning of a random quantum circuit.

\subsection{Results and Analysis}

\begin{figure}[H]
    \centering
    \includegraphics[width=\linewidth]{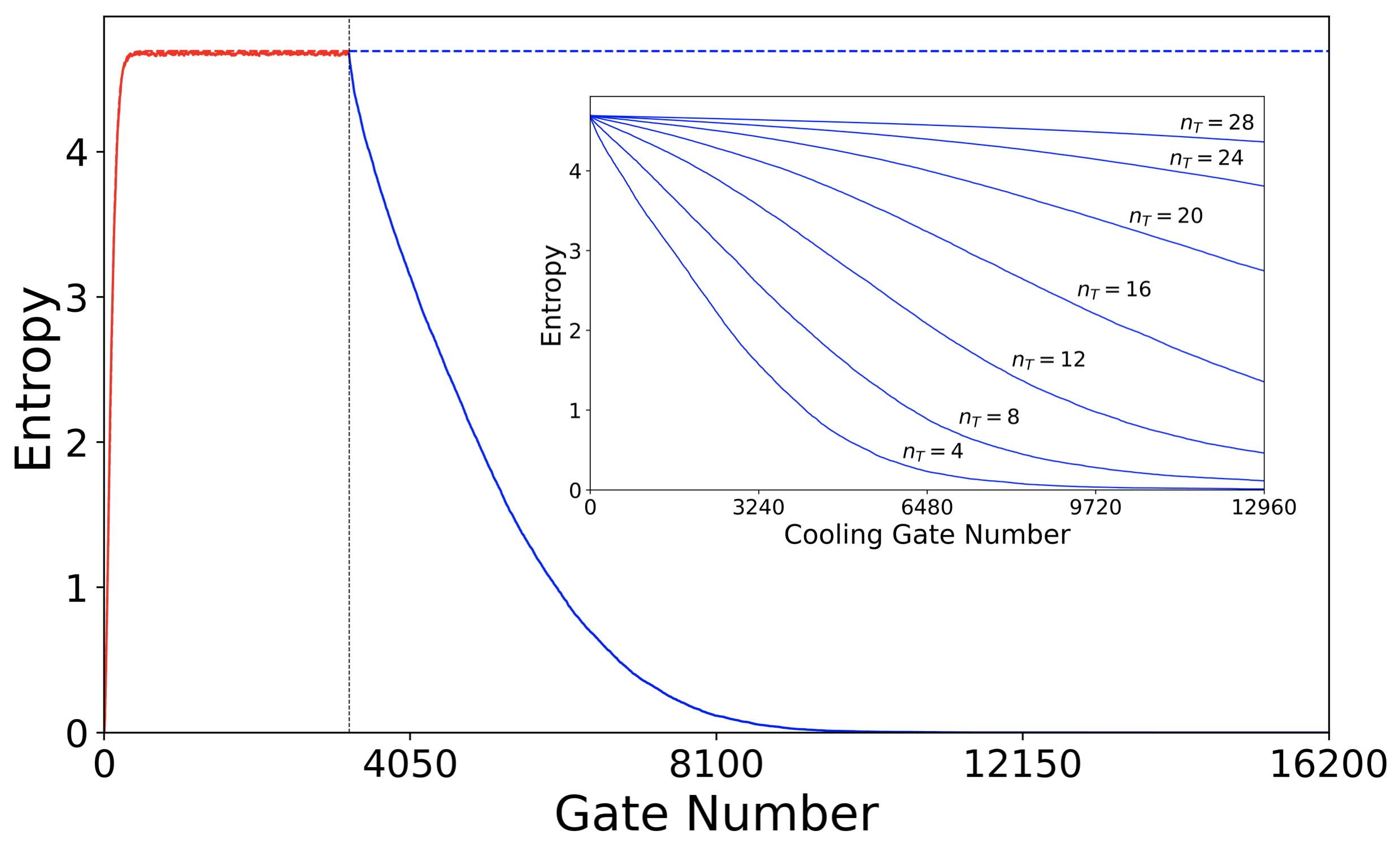}
    \caption{Average entanglement entropy of an $18-$qubit system vs. gate number of the applied heating (red) and cooling (blue) circuits over 200 realizations. \textit{Solid line:} Clifford entanglement cooling. \textit{Dashed line:} universal entanglement cooling. \textit{Inset:} Average entropy vs. gate number for multiple $18-$qubit Clifford${+}T$ circuits, this time showing only the cooling process, demonstrating a gradual transition from reversible (Clifford) to irreversible (universal) as the number of $T$ gates in the heating circuit increases.}
    \label{cooling}
\end{figure}

As expected \cite{chamon2014emergent, shaffer2014irreversibility}, the cooling algorithm succeeds in reversing all Clifford circuits $U^0$, reaching a factorized state in every realization. Conversely, the algorithm has no effect on the universal circuits, failing entirely to disentangle $U^U\ket{\psi_0}$ even a small amount; instead, the system remains at its maximum entropy during the entirety of the algorithm (Fig. \ref{cooling}). This behavior reiterates the role of the cooling algorithm in characterizing entanglement complexity as described above.

\begin{figure}[h!]
    \centering
    \includegraphics[width=\linewidth]{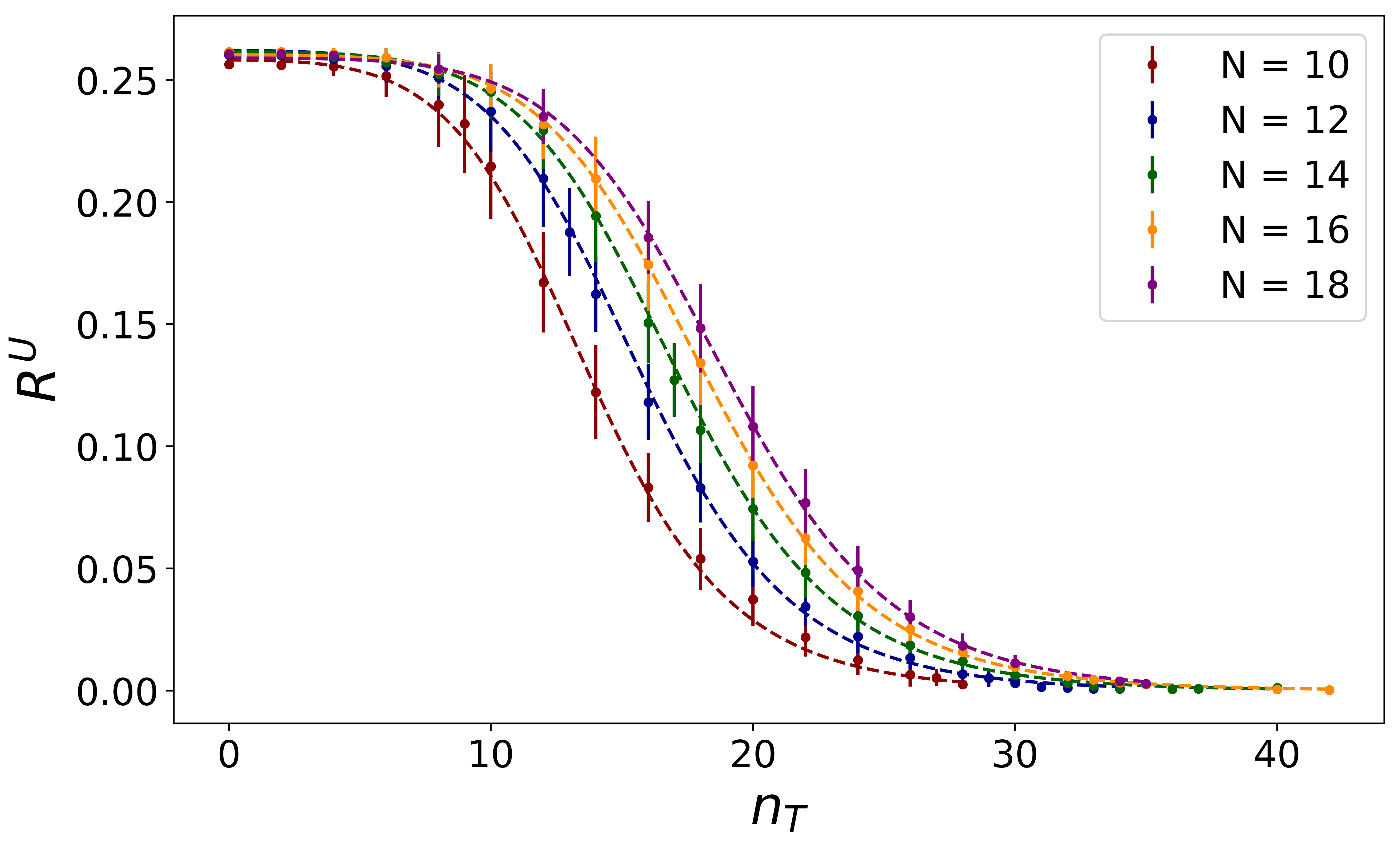}
    \caption{Reversibility $R^U$ (Eq. \eqref{eq2.4.1}) of Clifford$+T$ circuits $U^{n_T}$ as $n_T$ increases, fit by Eq. \eqref{eq3.3.1}. As with the analytical results of the variance $Var^U$ shown in Fig. \ref{var}, the reversibility of a random circuit can be seen approaching zero for circuits with high levels of doping.}
    \label{rev}
\end{figure}

Turning to the results from our Clifford${+}T$ simulations (Fig. \ref{rev}), we do in fact see confirmation of a gradual transition from non-complexity to complexity as $n_T$ increases and the algorithm's effectiveness lowers. Such a transition can be seen in the inset of Fig. \ref{cooling}; the more $T$ gates added to the heating circuit, the harder it becomes for the algorithm to lower the entropy of the system, until eventually the entropy cannot be lowered at all. It is important to note that trials involving only a small number of $T$ gates were \textit{sometimes} successful in disentangling $U\ket{\psi_0}$, with the likelihood of reaching a factorized state decreasing as $n_T$ increases until vanishing entirely for large $n_T$.

In analyzing the relationship between $n_T$ and the reversibility $R^U$ defined in Eq. \eqref{eq2.4.1}, we find the following fit of our data:
\be
\begin{split}
R^{U}&{=}
\\[-10pt]
&\frac{\gamma}{3.36\gamma^{1.04} + \exp\left[\mathrm{e}\sqrt{n_T}\right] \times 10^{-3} + 1.78} + \frac{0.2}{d^{1.25}}
\label{eq3.3.1}
\end{split}
\ee
where $\gamma = 4\log(d+560) - 109/3$. Here we see that a circuit $U$ does indeed become less reversible as each $T$ gate is added.

As stated above, low reversibility corresponds to an increase in entanglement complexity, eventually approaching universal complexity as $R^U$ approaches zero. Remarkably, we find that $R^U$ has the exact same lower bound of $0.2/d^{1.25}$ with increasing $n_T$ as $\overline{Var^U}$. We again calculate the minimum number of $T$ gates needed for Eq. \eqref{eq3.3.1} to approach this limit and fit the results:
\be
n_T^{min_R} = 0.7\left[2.29N-\frac{5}{3}\right]^{1.4} = 0.7 (n_T^{min_V})^{1.4}.
\label{eq3.3.2}
\ee

An important remark is that the algorithm uses a fixed number of accepted gates $40N^2$. This number is \textit{not} the total number of gates tried in the algorithm (i.e. does not include the number of operations that were rejected along the way). Additional trials are necessary to obtain an accepted gate when $n_T$ increases, and thus Eq. \eqref{eq3.3.1} is not a universal scaling function. The exact scaling is expected to depend on the parameters of the algorithm, such as the cooling schedule used. Nevertheless, the result above demonstrates that a circuit $U$ does indeed become less reversible as each $T$ gate is added.

Comparing all three calculated $T$ gate minima, we begin to paint a picture of the onset of the three characteristics of entanglement complexity analyzed in this work. The number of $T$ gates needed to induce this onset for a system of size $N$ is shown below in Table \ref{table1}, summarized for each characteristic. We can therefore see that small temporal entanglement fluctuations and complexity of entanglement are two aspects of the same thing, namely the entanglement complexity driven by the doping of non-Clifford gates that are spreading in the system.

{\renewcommand{\arraystretch}{2}
\begin{table}[H]
  \begin{center}
    {\setlength{\tabcolsep}{0.5em}
    \begin{tabular}{c|c|c}
      & {$n_T^{min}$} & {Order of Growth}\\
      \hline
      $\bm{D_{KL}}$ & $N+2$ & $\Theta(N)$\\
      \hline
      $\bm{\overline{Var^U}}$ & $2.29N-\frac{5}{3}$ & $\Theta(N)$\\
      \hline
      $\bm{R^U}$ & $0.7\left[2.29N-\frac{5}{3}\right]^{1.4}$ & $\Theta(N^{1.4})$\\
    \end{tabular}}
    \caption{Number of $T$ gates added to $U$ before crossing over into universal behavior of the ESS $(D_{KL})$, fluctuations $(\overline{Var^U})$, and reversibility $(R^U)$, as well as the order of growth of $n_T^{min}$ with respect to N.}
    \label{table1}
  \end{center}
\end{table}}

Additionally, one can draw a direct connection between the results put forth here and recent studies of barren plateaus in quantum machine learning \cite{McClean2018barren, Holmes2021barren, Cerezo2021barren, garcia2022barren}. In particular, the complexity transitions explored in this work reveal that the flattening of our training landscape (i.e. small entropy fluctuations brought on by adding non-stabilizer resources) significantly hinders the learnability of a successful disentangling circuit. The relationship between magic and barren plateaus will therefore be an important topic for future studies.

\section{Conclusions}

In this work, we study the transition in entanglement complexity in random Clifford circuits driven by the doping of local non-Clifford $T$ gates. Operator spreading by Clifford blocks applied after each $T$ gate results in the emergence of universal entanglement spectrum statistics and temporal fluctuations, typically associated with the onset of quantum chaos \cite{leone2021quantum, oliviero2021transitions}. In addition, this transition directly affects the performance of the entanglement cooling algorithm, a process which can be used to learn a disentangling circuit that will reverse the initial stochastic evolution. Irreversibility in a quantum circuit thus gradually emerges as entanglement complexity increases.

These results put forth additional questions regarding chaos characterization, such as what impact (if any) our choice of doping gate has on this complexity transition. If, for instance, we instead simulate Clifford + $T\! o\! f\! f\! oli$ circuits, how quickly would we reach a chaotic state? Additionally, other potential cost functions for the cooling algorithm may be explored in future works. The 2-Rényi stabilizer entropy, for instance, can also be calculated to detect the presence of chaos in quantum states \cite{leone2022stabilizer, campbell2011catalysis, goto2021chaos}. From the point of view of the theory of the Markov chain, it would be a decisive step forward to be able to prove why large fluctuations of the cost function ensure a greater performance of the entanglement cooling algorithm.

This work shows that entanglement complexity arises from the interplay of magic and entanglement, which has also been shown \cite{leone2021quantum} to be at the root of the onset of complex behavior as revealed by OTOCs \cite{harrow2020separation, roberts2017chaos}. We  believe that the same mechanism is responsible for a more general transition to complex quantum behavior, for instance in the retrieval of information (or lack thereof) from a black hole \cite{leone2022information}. Since the undoing of entanglement is equivalent to the learning of an unknown quantum circuit, the complexity of entanglement also has consequences for the complexity of quantum machine learning algorithms, which is the scope of further investigations.

Even more importantly, at the theoretical level, we see that neither the non-stabilizerness (induced by the $T$ gates) nor entanglement (due to the two-qubit Clifford gates) is on its own responsible for the onset of complex, universal entanglement. It is the marriage of the two -- that is, the operator spreading of non-Clifford resources -- that is responsible for the onset of complex quantum behavior. While this transition can be marked in terms of the adherence to universal ESS or the eight-point out-of-time-order correlation functions, one needs a more complete theoretical framework to describe the spreading of such magical resources and what makes a quantum system truly complex.

\section*{Acknowledgments}

We acknowledge support from NSF award number 2014000.

%%%%%%%%%%%%%%%%%%%%%%%%%%%%%%%%%%%%%%%%%%%%%%%%%%%%%%%%%%%%%%%%%%%%%%%
\bibliographystyle{bibliost}
\bibliography{biblio.bib}

\end{document}